\documentclass[%
reprint,
showpacs,preprintnumbers,
amsfonts,amsmath,amssymb,
aps,
prl,
]{revtex4-1}

\usepackage{epsfig}
\usepackage{mathrsfs}
\usepackage{amssymb}
\usepackage{amsbsy}
\usepackage{color}
\usepackage{cancel}
\usepackage[tight]{subfigure}

\usepackage{graphicx}
\usepackage[tight]{subfigure}
\usepackage{dcolumn}
\usepackage{bm}
\usepackage[colorlinks,linkcolor={blue},citecolor={blue}]{hyperref}

\newcommand{\ba}{\begin{array}}
\newcommand{\ea}{\end{array}}

\newcommand{\ra}{\rangle}
\newcommand{\la}{\langle}
\newcommand{\ci}[2][\sigma]{c_{#2#1}}
\newcommand{\cidag}[2][\sigma]{c_{#2#1}^{\dag}}

\newcommand{\fiop}[2][\sigma]{f_{#2#1}}
\newcommand{\fidag}[2][\sigma]{f_{#2#1}^{\dag}}

\newcommand{\eps}{\epsilon}

\newcommand{\up}{\uparrow}
\newcommand{\dn}{\downarrow}

\newcommand{\ket}[1]{\left\lvert#1\right\rangle}
\newcommand{\bra}[1]{\left\langle#1\right\rvert}

\newcommand{\half}{\frac{1}{2}}

\newcommand{\Fig}[1]{Fig.~\ref{#1}}

\begin{document}

\title{Synchronous and Asynchronous Mott Transitions in Topological Insulator Ribbons}

\author{Amal Medhi}\email{amedhi@physics.iisc.ernet.in} 
\author{Vijay B.~Shenoy}\email{shenoy@physics.iisc.ernet.in}
\author{H.~R.~Krishnamurthy}\email{hrkrish@physics.iisc.ernet.in} 
\affiliation{Center for Condensed Matter Theory, Indian Institute of Science, Bangalore 560012, India}


\begin{abstract}
We address how the nature of linearly dispersing edge states of two
dimensional (2D) topological insulators evolves with increasing
electron-electron correlation engendered by a Hubbard like on-site
repulsion $U$ in finite ribbons of two models of
topological band insulators. Using an inhomogeneous cluster slave rotor mean-field
method developed here, we show that electronic correlations drive the
topologically nontrivial phase into a Mott insulating phase via two
different routes. In a synchronous transition, the entire ribbon attains a Mott 
insulating state at one critical $U$ that depends weakly on the width of the ribbon. 
In the second, asynchronous route, Mott localization first occurs on the
edge layers at a smaller critical value of electronic interaction which then 
propagates into the bulk as $U$ is further increased until all layers of 
the ribbon become Mott localized. We show that the kind of Mott transition that takes
place is determined by certain properties of the linearly dispersing edge states which  characterize the topological resilience to Mott localization.

\end{abstract}

\maketitle

Topological insulators (TI) are a new quantum phase of matter
distinguished by a nontrivial topology of their electronic
state\cite{KanePRL05a,KanePRL05b,BernevigSCI06,BernevigPRL06,KonigJPSJ08,MooreNAT10,HasanRMP10}. 
The TI phase has been predicted and discovered in
numerous compounds starting from the two-dimensional quantum spin hall (QSH) 
systems\cite{BernevigSCI06, BernevigPRL06} to various three-dimensional
materials\cite{HsiehNAT08,HorPRB09,HsiehNAT09,ChenSCI09,HsiehPRL09}.
While the search is on for new materials with exotic topological
character\cite{ChadovNMAT10, LinNMAT10}, the phenomenon presents
opportunities to unravel new physics.

While the essential features of topological insulators (TI) are
manifestations of one-electron physics, there is intense interest to
explore the effect of electronic correlation and disorder on the
topological phase\cite{PesinNPHY10,RachelPRB10,VarneyPRB10,
YuPRL11,HePRB11,YamajiPRB11,LiPRL09}.
In TIs, strong spin-orbit coupling (SOC) leads to a
time-reversal invariant band structure with a bulk charge gap and
gapless edge modes which show Dirac like linear dispersion with a
characteristic velocity\cite{HasanRMP10}. The gapless edge states are
immune to weak perturbations (disorder/interactions) that are time
reversal symmetric.  A natural question that arises is regarding the fate
of the Dirac dispersion with increasing electronic
interactions/correlation.  The system is expected to evolve to a Mott
insulating state with increasing local repulsion characterized by a scale
$U$. What is the mechanisim of such a Mott transition, in particular,
does the edge mode velocity renormalize to zero, in an analogous
fashion as the Brinkman-Rice\cite{Brinkman1970} mechanisim of a
diverging effective mass in the Hubbard model?  Previous
works\cite{PesinNPHY10,RachelPRB10,VarneyPRB10,YuPRL11,HePRB11,YamajiPRB11}
which addressed similar issues revealed that the gapless surface states do survive
weak to moderate correlation while under stronger correlations the TI
phase evolves into various insulating phases e.g., a spin
liquid phase with gapped charge spectrum but gapless surface spinon
excitation or a bulk antiferromagnetic insulator etc.

Of particular interest is the role of the local repulsion $U$ on the
nature of electronic states in {\em finite systems} of TIs with
terminating edges (we focus on 2D in this paper), such as, for
example, a ribbon which is long along the $x$-direction but of
finite width $L$ along the $y$-direction.  The question of
how interaction affects differentially the gapless edge states and the
gapped bulk states existing simultaneously in such system of TIs with
finite boundaries is unexplored hitherto. Such a study requires a
detailed treatment that captures inhomogeneous nature of the
electronic state due to the lack of translational symmetry
perpendicular to the boundaries. This, along with the
necessity to tackle strong interactions makes the problem difficult,
even prohibitively expensive, to treat with accurate methods such as
exact numerical diagonalization \cite{VarneyPRB10}.  Here we study
interaction effects in TI ribbons by employing an {\em inhomogeneous}
cluster slave rotor mean-field (SRMF)
method\citep{FlorensPRB02,FlorensPRB04,ZhaoPRB07} which is known to
provide a correct qualitative description of the Mott
physics\cite{ZhaoPRB07,PesinNPHY10} while allowing for the treatment
of large system sizes. The effect of local interaction is captured by
introducing a Hubbard like on-site repulsion ($U$) into two well known
models of 2D TIs - the Kane-Mele
(KM)\cite{KanePRL05a,KanePRL05b} and Bernevig-Hughes-Zhang
(BHZ)\cite{BernevigSCI06} models. Our focus is on the dynamics in the
charge sector and the concomitant Mott localization engendered by
increasing $U$, following the spirit of ref.~\cite{Brinkman1970}.

The highlights of the study are the following.  With increasing $U$, a
finite ribbon of TI attains the Mott insulating state in {\em two
  distinct} ways. First, in the {\em synchronous} transition, charge
fluctuations vanish {\em simultaneously} on all layers (with different
$y$-coordinates) of the ribbon at a critical $U_{C}$
that depends on the ribbon width $L$. The second route to the 
Mott state is via an {\em asynchronous} transition. Here, the
charge fluctuations at the edge layers vanish at a critical 
$U_{E}$ which is independent of $L$. With further increase in $U$,
successive layers become Mott localized until at an $L$
dependent $U_{C}$ the entire ribbon is Mott insulating. 
A remarkable feature here is the simultaneous realization of two phases in the ribbon --
the Mott insulating (topologically trivial) state in outer layers and 
topologically non-trivial state in inner layers for $U_{E}<U< U_{C}$.
This results in gapless modes in the {\em bulk layers} that are at the
boundary between the Mott insulating and topologically non-trivial
regions. We discuss the physics underlying these results later in the paper.

\noindent
{\bf TI models:} We briefly describe the two models for topological 
insulators that we study here. The KM model\cite{KanePRL05a,KanePRL05b} 
describes electrons hopping on a honeycomb lattice,
\begin{align}
{\cal H}_{K} =& - t\sum_{\la i,j\ra \sigma}\cidag{i}\ci{j} 
+ i\lambda\sum_{\la\la i,j\ra\ra}
\sum_{\sigma\sigma'}\nu_{ij}\tau^z_{\sigma\sigma'}\cidag{i}\ci[\sigma']{j}
\label{km_model}
\end{align}
where $t$ is the nearest neighbor, $\la i,j\ra$ hopping amplitude and $\lambda$ is
spin-orbit coupling strength. $\la\la i,j\ra\ra$ denote next nearest neighbor
sites. $\tau^z$ is $z$-Pauli spin matrix and $\nu_{ij}=\pm 1$
depending on the orientation of the hop\cite{KanePRL05a}. 
For $\lambda \ne 0$, the resulting dispersion has non-trivial topology. 
The BHZ model\cite{BernevigSCI06} defined on a square lattice with four spin-orbit coupled 
atomic orbitals, e.g., $\ket{s\up}$, $\ket{p\up} \equiv \ket{\left(p_y+ip_x\right)\up}$,
$\ket{s\dn}$, and $\ket{p\dn} \equiv \ket{\left(p_y-ip_x\right)\dn}$ is,
\begin{align}
{\cal H}_{B} = \sum_{i\alpha\sigma}\eps_\alpha\cidag{i\alpha}\ci{i\alpha} - \sum_{i\bm{\delta}\alpha\beta\sigma}
t_{\bm{\delta}\sigma,\alpha\beta} \cidag{i\alpha}\ci{i+\bm{\delta}\beta}
\label{eq:bhzmodel}
\end{align}
where $\alpha, \beta = s, p$ and $\sigma = \up, \dn$. $\eps_\alpha$ is the 
energy of the spin-orbit coupled orbital $\ket{\alpha\sigma}$ and $\bm{\delta}$ denote
a nearest neighbor vector. The hopping matrix elements 
$t_{\bm{\delta}\sigma,\alpha\beta}$ in the $\ket{s\sigma}$, $\ket{p\sigma}$ basis
is given by (see also \cite{FuPRB07}),
\begin{align}
t_{\pm\hat{x}\sigma} = \begin{pmatrix}
t_{ss} & \pm\sigma\frac{it_{sp}}{\sqrt{2}}  \\
\pm\sigma\frac{it_{sp}}{\sqrt{2}} & -t_{pp} \\
\end{pmatrix},
\;\;
t_{\pm\hat{y}\sigma} = \begin{pmatrix}
t_{ss} & \pm\frac{t_{sp}}{\sqrt{2}}\\
\mp\frac{t_{sp}}{\sqrt{2}} & -t_{pp}\\
\end{pmatrix}
\end{align}
where $t_{ss}$, $t_{sp}$, $t_{pp}$ are overlap integrals and $\sigma$ takes 
values $+1$ ($-1$) for spin $\up$ ($\dn$).
We set $t_{ss} = t_{pp} = t$ and $t_{sp}/\sqrt{2} = v/2$, and 
define $\eps_0$ such that $\eps_s= -(\eps_0-4t)$ and 
$\eps_p = (\eps_0-4t)$. Thus we are left with three parameters for the model -- 
$t$, $v$ and $\eps_0$. The model shows inverted band structure and 
hence the topological phase for $0 < \eps_0 < 8t$.

\noindent
{\bf Interactions:} We introduce electron interaction via an on site Hubbard repulsion,
\begin{align}
{\cal H}_U = \frac{U}{2}\sum_i n_i(n_i-1) \label{Hubbard}
\end{align}
to the two models in Eq.~(\ref{km_model}) and (\ref{eq:bhzmodel}) where $n_i$ 
is the electron number operator. For the KM model $n_i = \sum_{\sigma}\cidag{i}\ci{i}$, 
while for the BHZ it is $n_i = \sum_{\alpha\sigma}\cidag{i\alpha}\ci{i\alpha}$,

\noindent
{\bf Inhomogeneous SRMF Formulation:} We now outline the inhomogeneous cluster SRMF 
formulation developed here to study the above two interacting models for lattices 
with ribbon geometry. In the usual slave rotor (SR) 
formulation\citep{FlorensPRB02,FlorensPRB04,ZhaoPRB07}, the electron operator is expressed 
as a product of a fermionic spinon ($f$) which carries the spin
and a bosonic rotor $(\theta)$ which carries the charge. One writes
$\cidag{i\alpha} = \fidag{i\alpha}e^{-i\theta_i}$ subject to the constraints, 
$\sum_{\alpha\sigma} n_{i\alpha\sigma}^f + n_i^\theta = 1$ and 
$\sum_{\sigma} n_{i\sigma}^e = \sum_{\alpha\sigma} n_{i\alpha\sigma}^f$, where 
$f^{\dag}_{i\alpha\sigma}$ ($e^{-i\theta_i}$) is the spinon creation (rotor annihilation) operator.
The resulting Hamiltonian (${\cal H}_{SR}$) in terms of the auxiliary operators is then mean-field
decoupled as follows. Positing the ground state of ${\cal H}_{SR}$ to
be the direct product of the spinon and rotor ground states,
$\ket{\Psi} = \ket{\Psi_f} \ket{\Psi_{\theta}}$, one defines 
effective spinon and rotor Hamiltonians as ${\cal
  H}^f=\bra{\Psi_{\theta}}{\cal H}^{SR}\ket{\Psi_{\theta}}$ and ${\cal
  H}^{\theta}=\bra{\Psi_{f}}{\cal H}^{SR}\ket{\Psi_{f}}$,
respectively. For the BHZ model, this gives
\begin{align}
{\cal H}^f_{B} =& \sum_{i\alpha\sigma}(\epsilon_{\alpha}-\mu_f)\fidag{i\alpha}\fiop{i\alpha}
- \sum_{i\bm{\delta}\alpha\beta\sigma} t^{f} 
\fidag{i\alpha}\fiop{i+\bm{\delta}\beta} \label{eq:bhz_spinon_model} \\
{\cal H}^{\theta}_{B} =& - \sum_{i\bm{\delta}\alpha\beta\sigma} t^{\theta}
e^{-i\theta_{i}}e^{i\theta_{i+\bm{\delta}}}
+ \frac{U}{2}\sum_i\left(n_i^{\theta} - \mu_\theta \right) n_i^\theta\label{eq:bhz_rotor_model}
\end{align}
where chemical potentials are introduced to control the mean particle density. The effective 
hopping parameters $t^{f}$ and $t^{\theta}$ are given by
$
t^{f}_{i\bm{\delta}\sigma,\alpha\beta} = t_{\bm{\delta}\sigma,\alpha\beta} B_{i,i+\bm{\delta}}
$
and
$
t^{\theta}_{i\bm{\delta}\sigma,\alpha\beta} =
t_{\bm{\delta}\sigma,\alpha\beta}\chi_{i\alpha,i+\bm{\delta}\beta}^{\sigma}
$,
where $B_{ij} = \bra{\Psi_{\theta}} e^{-i(\theta_i-\theta_j)} \ket{\Psi_{\theta}}$ and 
$\chi_{i\alpha,j\beta}^{\sigma} = \bra{\Psi_{f}} f^{\dag}_{i\alpha\sigma}f_{j\beta\sigma} \ket{\Psi_{f}}$  
(analogous expressions are used for the KM model). The resulting spinon and rotor 
Hamiltonians are solved self consistently to obtain the ground state. 
A key point to be noted here is that the $B$s and $\chi$s are {\em bond dependent} 
in a ribbon, i.~e., they are inhomogeneous. The spinon Hamiltonian is quadratic 
albeit with inhomogeneous renormalized hopping parameters and hence can 
be diagonalized numerically\footnote{Since our focus is on the charge sector and the
associated Mott physics in the spirit of \cite{Brinkman1970}, we keep aside the magnetic correlations which are typically 
introduced by adding an exchange term in the Hamiltonian\cite{RachelPRB10} and 
treating it via mean field theory.}
The rotor problem, however, is non-quadratic and a full numerical solution 
is still prohibitive, and additional approximations such as cluster mean 
field method have to be adopted. In conventional cluster mean-field method\cite{ZhaoPRB07},
one considers a small cluster of sites where the rotor Hamiltonian is treated exactly. 
Terms connecting the sites inside the cluster to sites outside (bath) are decoupled by using,
$\Phi_i=\la e^{i\theta_i}\ra$
where $\Phi_i$ is a site-dependent  mean-field parameter also to be obtained self-consistently. This quantity
plays an important role in that it signifies charge fluctuation at the site $i$. A nonzero $\Phi_i$ 
implies charge fluctuation whereas vanishing of $\Phi_i$ implies a ``local'' Mott  insulating phase. 

\begin{figure}[htb]
 \centering
  \includegraphics[width=0.35\columnwidth]{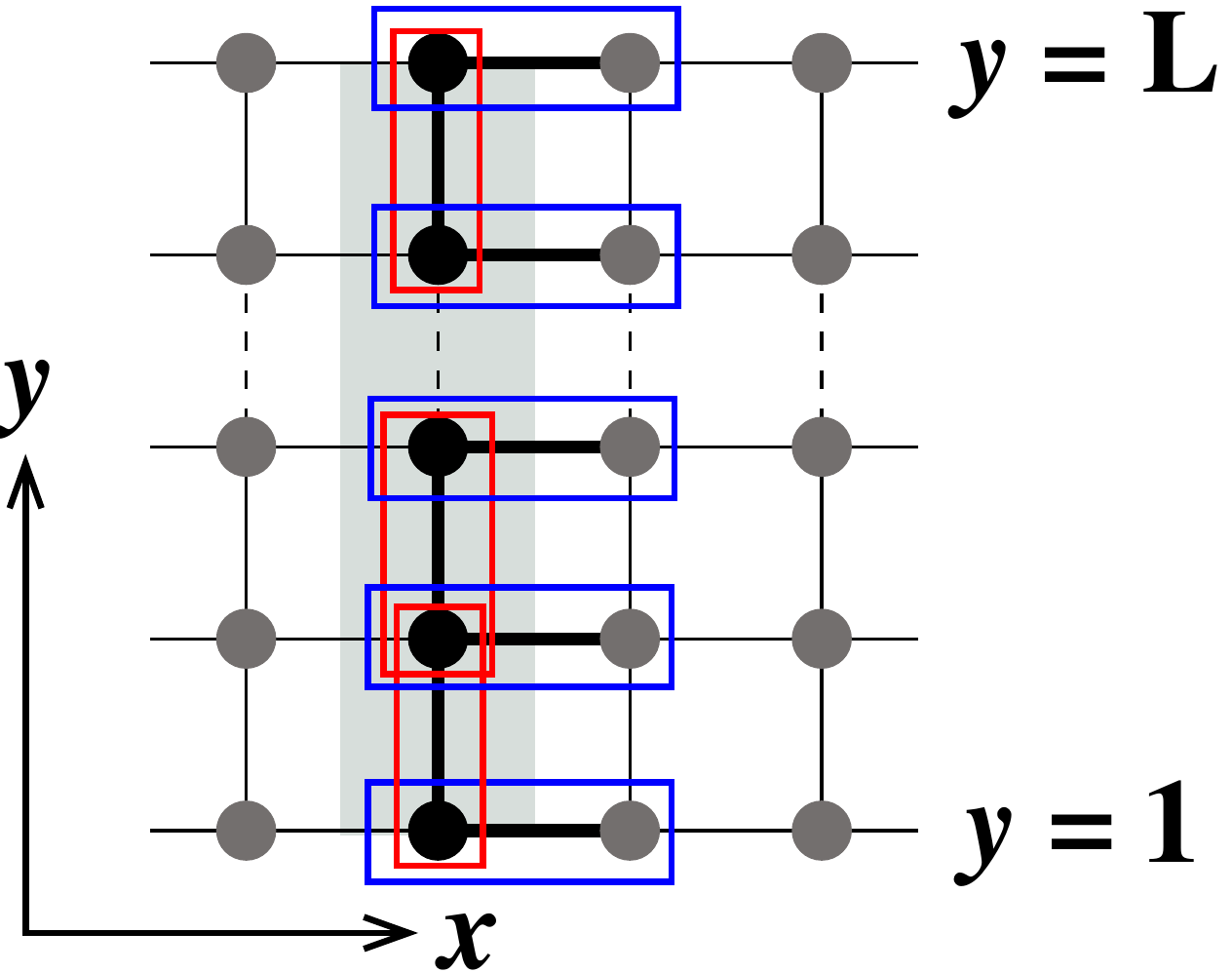}
  \caption{Inhomogeneous cluster mean-field method. Each unique bond in the strip which is
  treated as a distinct cluster is shown enclosed by a rectangle. The shaded region represents
  a unit cell of the strip.}
  \label{fig:cluster}
\end{figure}

To take into account the inhomogeneity introduced by the lost 
translational symmetry in the $y$-direction, we first express the ${\cal H}^{\theta}_B$ as
a sum of bond terms, i.e., as ${\cal H}^{\theta}_B = \sum_{b_{ij}} h^{\theta}_{ij}$.
Now we treat the bonds (such as those shown in \Fig{fig:cluster}) one at a time 
as our cluster and the rest as the bath and obtain a 
mean-field two-site cluster Hamiltonian for each bond. Each of the bond problem then 
solved to obtain a self-consistent set of solutions for 
$\{\Phi_y\},\; y=1,\ldots,L$. We treat all the unique
bonds (coloured boxes in \Fig{fig:cluster}) in the super-cell 
(indicated by the shaded region in \Fig{fig:cluster}) 
utilizing the translational symmetry in the $x$-direction. 
The above inhomogeneous calculation becomes numerically expensive (for
typical width of $L=100$ considered here) as it involves
diagonalization of a large number of rotor Hamiltonians in each
iteration of the self consistency loop.

We also obtain the bulk phase diagram (lattice with no boundaries) where the
parameters $\Phi$, $B$, $\chi$ are all site independent
(homogeneous). For this, it suffices to solve the rotor mean-field Hamiltonian
only for a single cluster. In the results discussed below all energies
are measured in units of appropriate $t$ both for the KM and BHZ
models.

\begin{figure}[htb]
\centerline{\includegraphics[width=0.45\columnwidth]{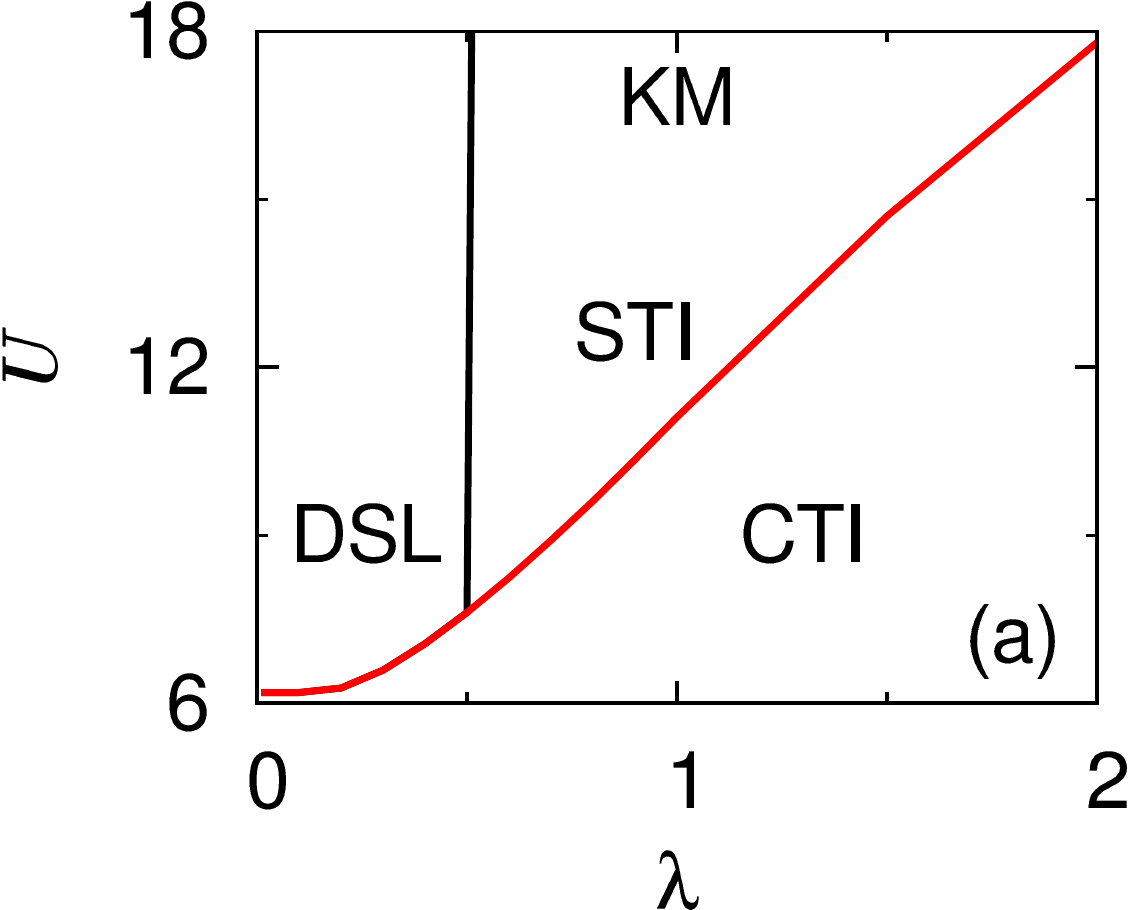}~~~\includegraphics[width=0.45\columnwidth]{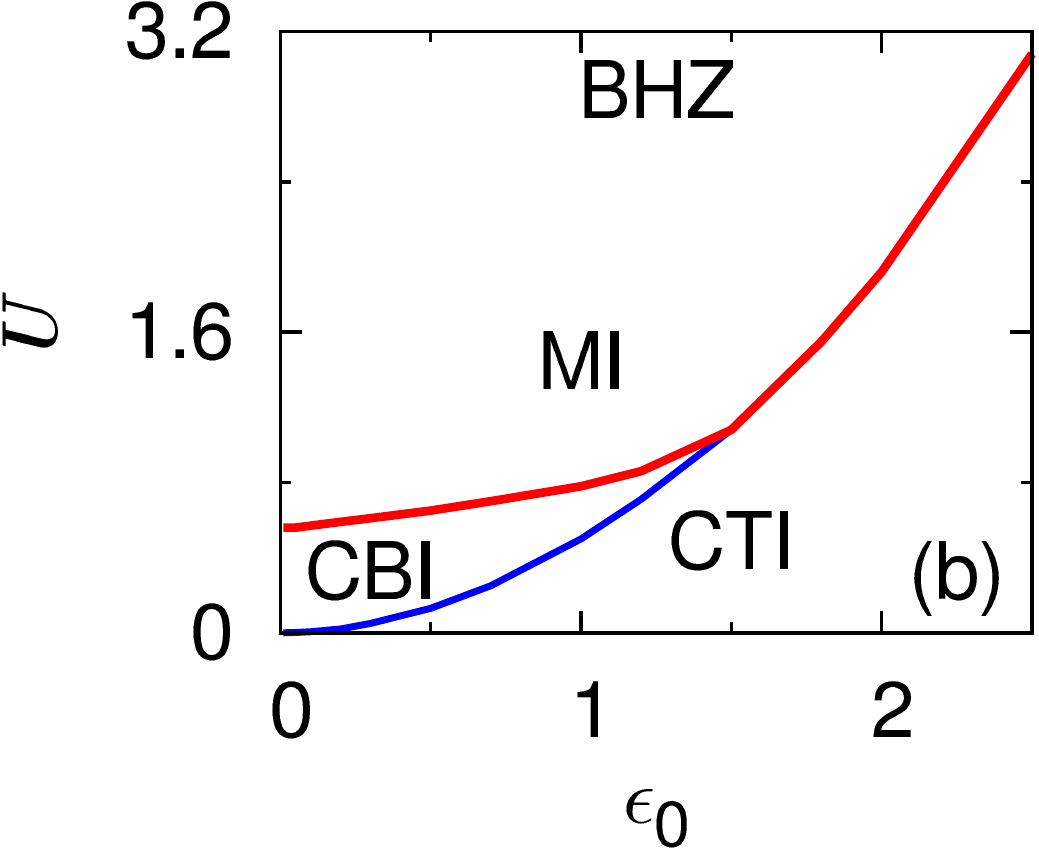}}
  \caption{Bulk phase diagram - (a) KM model: There are three phases:- Correlated Topological Insulator (CTI), DSL (Dirac Spin Liquid), STI (Spinon Topological Insulator) phases. DSL/STI phases are Mott insulating. (b) BHZ model: There are three phases:- CTI, Correlated Band Insulator (CBI) and MI (Mott insulator). }
  \label{fig:bulk_phasediagram}
\end{figure}

{\bf KM Model Results:}  We first discuss the bulk phase diagram of the KM model 
(see \Fig{fig:bulk_phasediagram}(a)) obtained within our formulation. For any 
given $\lambda$ there is a critical $U_B$ such for $U > U_B$ a Mott insulating state is obtained. 
For $U< U_B$ (region below the red line \Fig{fig:bulk_phasediagram}(a)), there are local 
charge fluctuations even though there is a charge/spin gap, and the spinon dispersion is 
topologically nontrivial -- we call this state a correlated topological insulator (CTI). 
There are two regimes of $\lambda$ for which the Hubbard interaction produces different 
types of Mott states. For $\lambda < \lambda_c = \half$, the Mott state is a Dirac spin 
liquid (DSL), i.~e., a state with no charge fluctuations and a gapless spinon dispersion similar 
to that of graphene. In this regime effective $\lambda$ for the spinon hopping renormalizes 
to zero. For $\lambda > \lambda_c$, the Mott state has a gapped topologically 
non-trivial spinon spectrum (spinon topological insulator (STI)) - the state has local spin 
fluctuations but no charge fluctuations. 
All the boundaries in \Fig{fig:bulk_phasediagram} correspond to second order 
transitions in that $\Phi$ goes to zero continuously with increasing $U$.

\begin{figure}[htb]
 \centering
  \includegraphics[width=0.48\columnwidth]{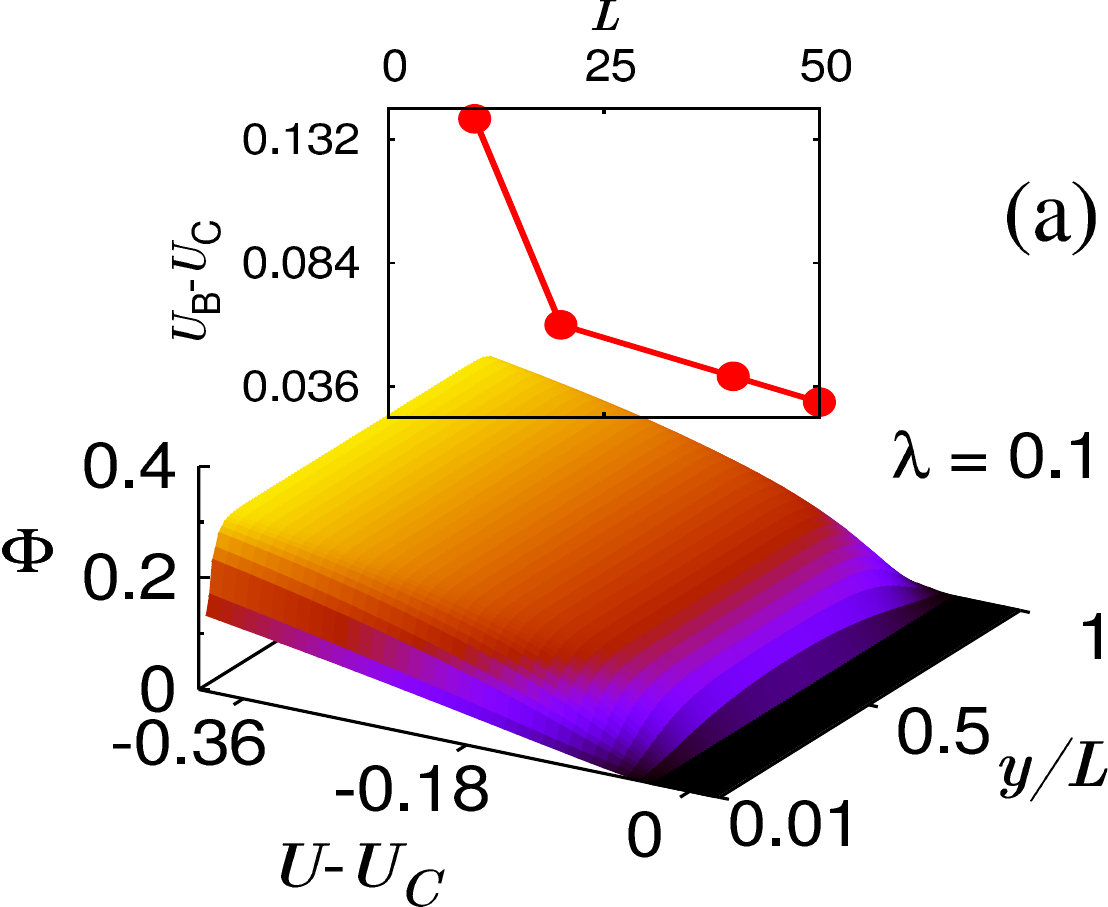}
  \hfill
  \includegraphics[width=0.48\columnwidth]{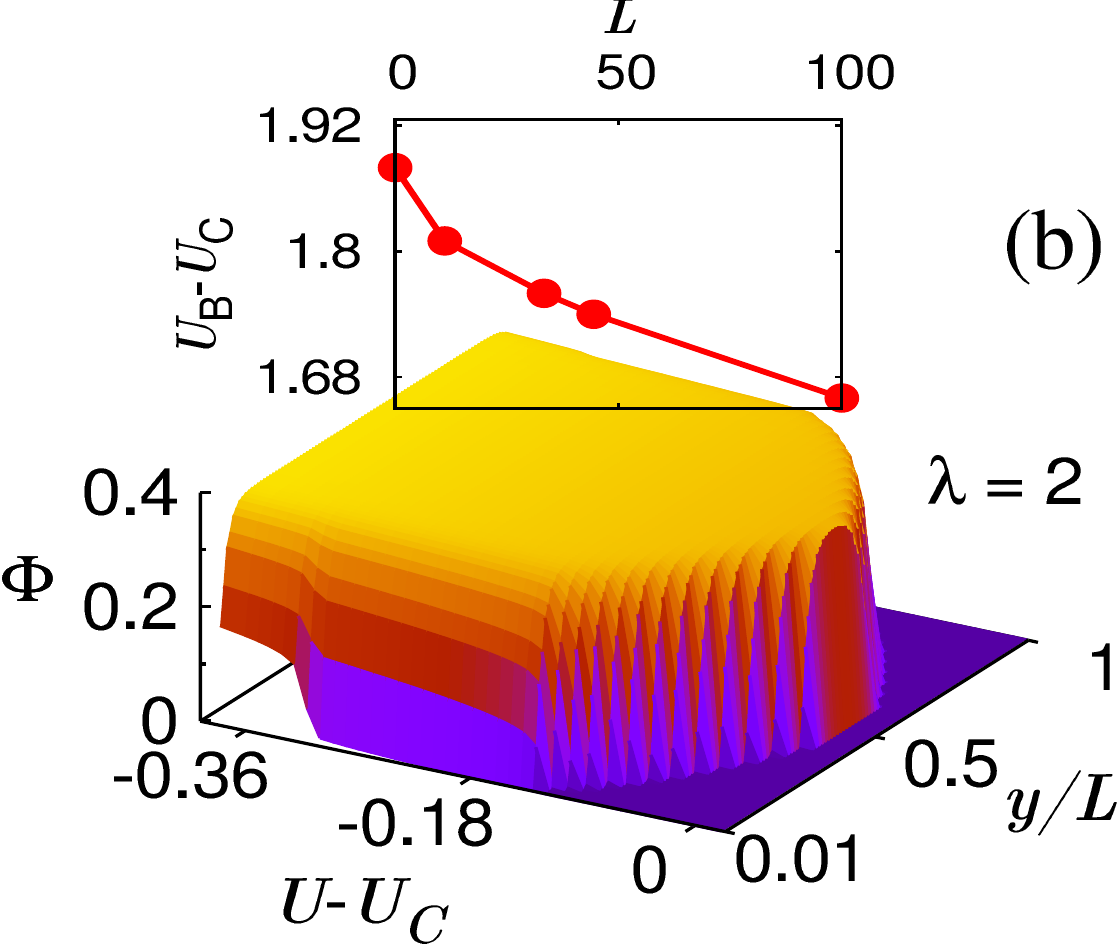}
  \caption{Mott transition in KM ribbons of width $L=50$. in the two different regimes of $\lambda$. (a) Synchronous Mott transition obtained for $\lambda = 0.1$ where the entire ribbon becomes Mott insulating at a critical value of $U_C$. The bulk value of critical Hubbard interaction is $U_B = 6.19$. (b) Asynchronous Mott transition for $\lambda = 2$ ($U_B =17.8$). The edge sites go Mott insulating at $U_{E} = 15.8$, followed by successive layers becoming Mott insulating with increasing $U$ until the ribbon becomes fully Mott insulating at $U=U_C$. Dependence of $U_C$ on the width $L$ of the ribbon is shown in the upper insets in both cases.  }
  \label{fig:km_mott}
\end{figure}

We now turn to zigzag edge terminated ribbons of finite
width. \Fig{fig:km_mott} shows plot of $\Phi_y$ as a function of $U$
and $y$ for two values of $\lambda$, e.g., $0.1$ and $2$. 
For $\lambda = 0.1 < \lambda_c$, the bulk value of the critical
Hubbard interaction is $U_B = 6.19$. As shown in \Fig{fig:km_mott}(a),
for this value of $\lambda$, the Mott transition of the ribbons occurs
in a ``{\em synchronous}'' fashion, i.~e., the charge fluctuations vanish
throughout the strip at a critical value of $U_C$. $U_C$ is quite
close to $U_B$ and the difference between $U_B$ and $U_C$ falls with
increasing width of the ribbon (see inset in \Fig{fig:km_mott}(a)).

For the value of $\lambda = 2 > \lambda_c$ we find completely
different physics. In this case, the edge sites (see
\Fig{fig:km_mott}(b)) undergo a local Mott transition at a value of
$U_{E} = 15.8$ which is significantly smaller than the bulk value for
the Mott transition $U_B = 17.8$ while the sites in the bulk continue
to enjoy charge fluctuations. Most interestingly, this critical value
$U_{E}$ does not depend on the with $L$ of the ribbon over the range
of ribbon widths $L=10$ to $100$ studied in this work. Further increase
of $U$ above $U_E$ results in successive layers attaining the Mott
insulating state (\Fig{fig:km_mott}(b)) -- a phenomenon we call 
``{\em asynchronous}'' Mott transition. The process of asynchronous Mott
transition continues until a width dependent critical $U_{C}$ is
attained. $U_{C}$ is always less than $U_{B}$ (see inset of
\Fig{fig:km_mott}(b)) and $U_{C} \rightarrow U_{B}$ as $L$ becomes
larger. 

The key point to be noted is that the nature of the Mott transition in
KM ribbons is determined by $\lambda$. For all $\lambda < \lambda_c$
we find synchronous Mott transition, while the second regime $\lambda
> \lambda_c$, asynchronous behaviour is obtained.

\begin{figure}[htb]
 \centering
  \centerline{ \includegraphics[width=0.48\columnwidth]{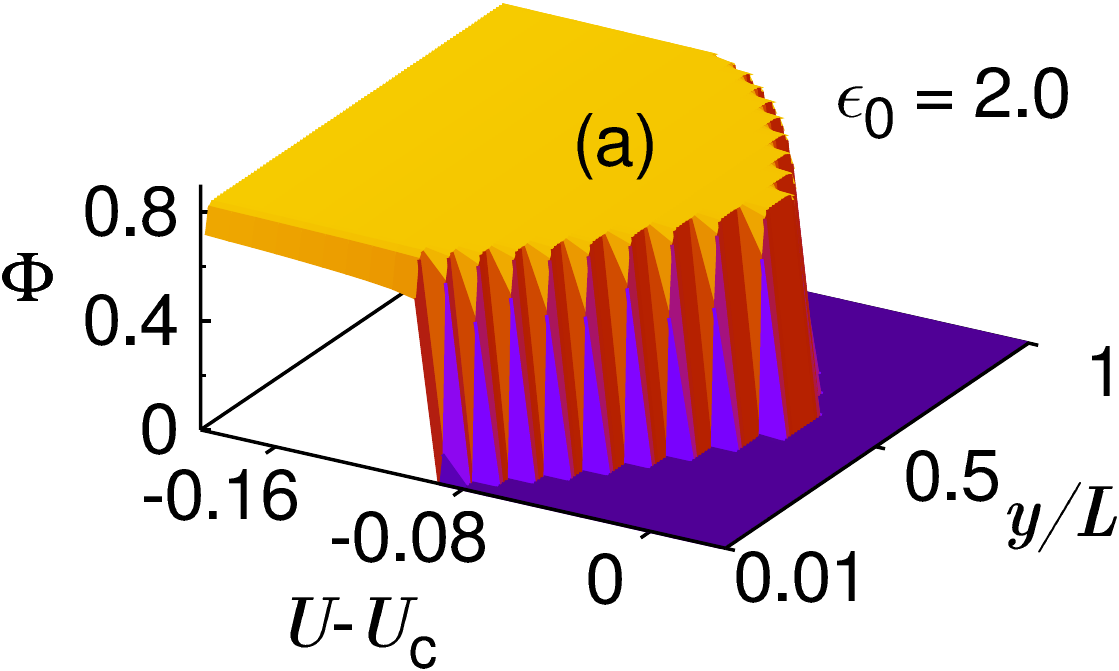}~~  \includegraphics[width=0.48\columnwidth]{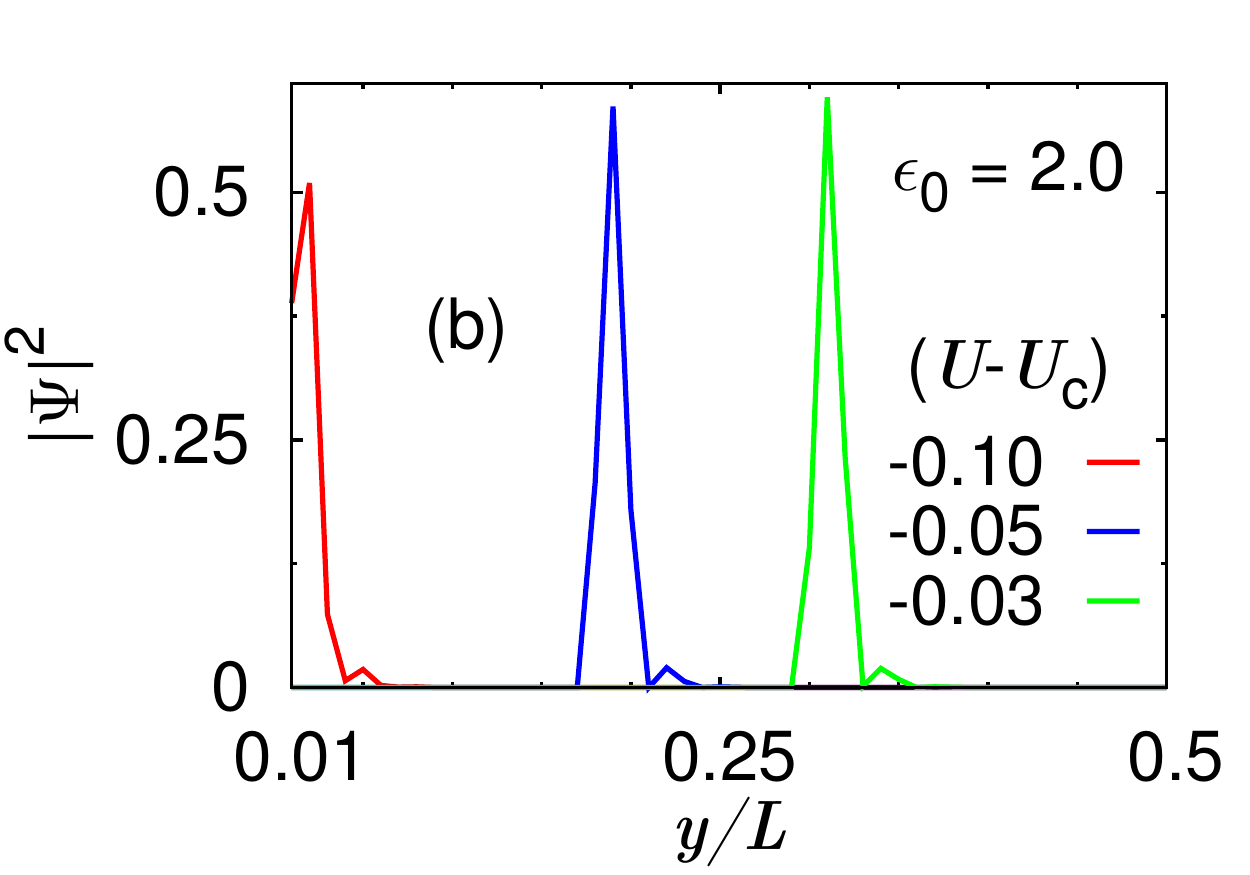}}
  \caption{{\bf (a)} Asynchronous Mott transition in a BHZ ribbon with $L=100$. {\bf (b)} Wave function of mode that appears at the MI -- CTI interface in the
asynchronous Mott transition. For $U \le U_E$, the mode is localized
at the edge of the sample (red curve). With increasing $U > U_E$ the
outer layers become Mott insulating and the interface modes appear in
the inside layers of the ribbon.
}
  \label{fig:bhz_phi}
\end{figure}

{\bf BHZ Model Results:} The results discussed here are for $t=v=1$. \Fig{fig:bulk_phasediagram}(b) shows the bulk phase diagram of the BHZ model.  There are again two regimes of the topological parameter. For $\eps_0 < 3/2$, there are two transitions with increasing $U$. In this regime first transition is at $U_{A}$, where the spinon dispersion becomes topologically trivial, but the system is still not Mott insulating, i.~e., the system effectively is a correlated band insulator (CBI).
At a larger critical value  $U_{B}$, the CBI state undergoes a Mott transition and obtains paramagnetic Mott insulator. The spinon dispersion renormalizes to zero. When $\eps_0 > 3/2$, the CTI phase gives way to the MI phase without the intervening CBI phase. The Mott transition in this case is a first order transition (unlike in the KM case).

\Fig{fig:bhz_phi}(a) shows asynchronous Mott transition in a BHZ
ribbon for $\eps_0 =2$. In fact, all BHZ ribbons that we have studied
over a range of $\eps_0$ and widths $L$ undergo asynchronous
transitions. All other qualitative features of the transition are
similar to that of the KM model. An interesting aspect is that in both
models, the asynchronous Mott transition leads to a state with trivial
topology. Thus between $U_E$ and $U_C$ the ribbon consists of two
regions, one adjoining the edges that is Mott insulating and
topologically trivial, while the center portion continues to be a CTI
endowed with a non-trivial topology. Thus one expects interface modes
to appear at the layer that separates the two regions which is now in
the bulk of the ribbon and to evolve further into the bulk as $U$
increases from $U_E$ to $U_C$. Indeed, we do find such spinon modes as
shown for the BHZ model in \Fig{fig:bhz_phi}(b). KM model also has
similar physics in the asynchronous regime.

\begin{figure}[htb]
 \centering
  \includegraphics[width=0.48\columnwidth]{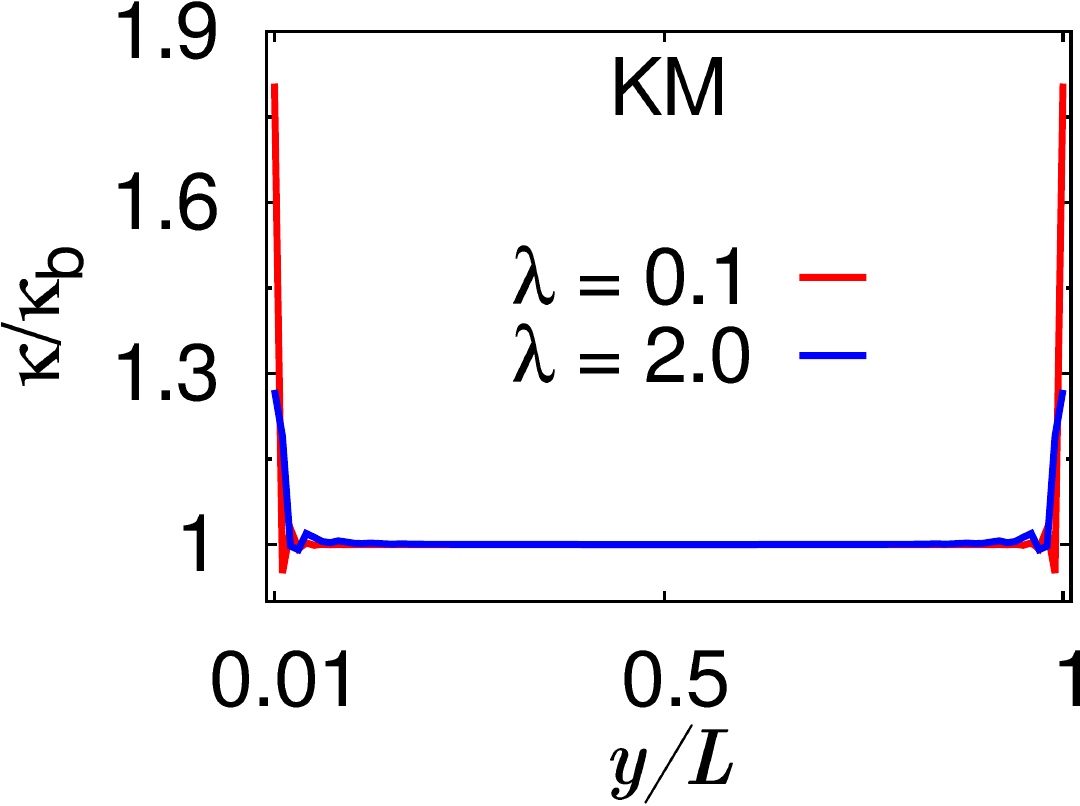}
  \caption{Local site compressibility $\kappa(y)/\kappa_B$ for the KM model.}
  \label{fig:compressibility}
\end{figure}

Why are there two types of transitions and what governs this? The
nature of the transition is governed by the local nature of the
topological edge states. The quantity $\kappa(y) = \lim_{\beta
  \rightarrow \infty} \int_0^\beta d\tau \langle n_y(\tau)n_y(0)
\rangle - \langle n_y(0)\rangle^2$, where $\beta$ is inverse temperature, 
  calculated using the non-interacting
  Hamiltonian is a measure of the local compressibility of a site in a
  layer at a given $y$. If the edge states brought about by the
  topological dispersion results in a large compressibility compared
  to the bulk, then the edge sites resist Mott localization --
  topological resilience to Mott localization -- since much
  kinetic energy is lost in localizing carriers at the edges.  Indeed for the KM model,
  the velocity of edge states for $\lambda \ll \lambda_c$ is $6
  \lambda$ while for $\lambda = 2$ the edge state velocity is $0.75
  \lambda$. This is reflected in a large site compressibility at the
  edges compared to the bulk value ($\kappa_B$) for $\lambda = 0.1$ (see
  \Fig{fig:compressibility}). For highly compressible edges, Mott
  transition of the ribbons occurs at value of $U$ close to that of
  the bulk value and occurs in a synchronous fashion. In the other
  limit where the edge site compressibility is comparable or lower to
  that of the bulk (as is the case for $\lambda =2$ in KM, and for the
  BHZ model), Hubbard energy dominates since there is no significant
  kinetic energy to be gained and Mott localization occurs at value of
  $U$ significantly lower than the bulk. The kinetic energy of the
  bulk spinon modes comes to play; these modes prevent the synchronous
  Mott transition and rendering it asynchronous. Our arguments, if
  applied, to the usual one band square lattice half-filled Hubbard
  model (no edge states) will predict an asynchronous transition for
  finite ribbons -- indeed we do find this in our SRMFT simulations of
  such systems.

Interestingly, for the synchronous Mott transition the spinon edge
mode velocity is continuously renormalized to lower values as $U$
approaches $U_C$. In the asynchronous case, the edge modes ``switch
layers'' and eventually vanish at $U_C$. This provides a clear
physical picture of the nature of the Mott transition in finite
ribbons of TIs.  Another important outcome of this study pertains to the use
of surface probes to investigate the electronic state correlated
materials (particularly topological insulators). A key point to
be borne in mind is that surface may show insulating character while
the bulk may still be locally compressible.
 
VBS thanks DST (Ramanujan grant) and DAE (SRC grant),  HRK thanks DST (Bose grant) for support.

\bibliography{bibliography_topins}

\begin{thebibliography}{27}%
\makeatletter
\providecommand \@ifxundefined [1]{%
 \@ifx{#1\undefined}
}%
\providecommand \@ifnum [1]{%
 \ifnum #1\expandafter \@firstoftwo
 \else \expandafter \@secondoftwo
 \fi
}%
\providecommand \@ifx [1]{%
 \ifx #1\expandafter \@firstoftwo
 \else \expandafter \@secondoftwo
 \fi
}%
\providecommand \natexlab [1]{#1}%
\providecommand \enquote  [1]{``#1''}%
\providecommand \bibnamefont  [1]{#1}%
\providecommand \bibfnamefont [1]{#1}%
\providecommand \citenamefont [1]{#1}%
\providecommand \href@noop [0]{\@secondoftwo}%
\providecommand \href [0]{\begingroup \@sanitize@url \@href}%
\providecommand \@href[1]{\@@startlink{#1}\@@href}%
\providecommand \@@href[1]{\endgroup#1\@@endlink}%
\providecommand \@sanitize@url [0]{\catcode `\\12\catcode `\$12\catcode
  `\&12\catcode `\#12\catcode `\^12\catcode `\_12\catcode `\%12\relax}%
\providecommand \@@startlink[1]{}%
\providecommand \@@endlink[0]{}%
\providecommand \url  [0]{\begingroup\@sanitize@url \@url }%
\providecommand \@url [1]{\endgroup\@href {#1}{\urlprefix }}%
\providecommand \urlprefix  [0]{URL }%
\providecommand \Eprint [0]{\href }%
\providecommand \doibase [0]{http://dx.doi.org/}%
\providecommand \selectlanguage [0]{\@gobble}%
\providecommand \bibinfo  [0]{\@secondoftwo}%
\providecommand \bibfield  [0]{\@secondoftwo}%
\providecommand \translation [1]{[#1]}%
\providecommand \BibitemOpen [0]{}%
\providecommand \bibitemStop [0]{}%
\providecommand \bibitemNoStop [0]{.\EOS\space}%
\providecommand \EOS [0]{\spacefactor3000\relax}%
\providecommand \BibitemShut  [1]{\csname bibitem#1\endcsname}%
\let\auto@bib@innerbib\@empty
\bibitem [{\citenamefont {Kane}\ and\ \citenamefont
  {Mele}(2005{\natexlab{a}})}]{KanePRL05a}%
  \BibitemOpen
  \bibfield  {author} {\bibinfo {author} {\bibfnamefont {C.~L.}\ \bibnamefont
  {Kane}}\ and\ \bibinfo {author} {\bibfnamefont {E.~J.}\ \bibnamefont
  {Mele}},\ }\href@noop {} {\bibfield  {journal} {\bibinfo  {journal} {Phys.
  Rev. Lett.}\ }\textbf {\bibinfo {volume} {95}},\ \bibinfo {pages} {146802}
  (\bibinfo {year} {2005}{\natexlab{a}})}\BibitemShut {NoStop}%
\bibitem [{\citenamefont {Kane}\ and\ \citenamefont
  {Mele}(2005{\natexlab{b}})}]{KanePRL05b}%
  \BibitemOpen
  \bibfield  {author} {\bibinfo {author} {\bibfnamefont {C.~L.}\ \bibnamefont
  {Kane}}\ and\ \bibinfo {author} {\bibfnamefont {E.~J.}\ \bibnamefont
  {Mele}},\ }\href@noop {} {\bibfield  {journal} {\bibinfo  {journal} {Phys.
  Rev. Lett.}\ }\textbf {\bibinfo {volume} {95}},\ \bibinfo {pages} {226801}
  (\bibinfo {year} {2005}{\natexlab{b}})}\BibitemShut {NoStop}%
\bibitem [{\citenamefont {Bernevig}\ \emph {et~al.}(2006)\citenamefont
  {Bernevig}, \citenamefont {Hughes},\ and\ \citenamefont
  {Zhang}}]{BernevigSCI06}%
  \BibitemOpen
  \bibfield  {author} {\bibinfo {author} {\bibfnamefont {B.~A.}\ \bibnamefont
  {Bernevig}}, \bibinfo {author} {\bibfnamefont {T.~L.}\ \bibnamefont
  {Hughes}}, \ and\ \bibinfo {author} {\bibfnamefont {S.-C.}\ \bibnamefont
  {Zhang}},\ }\href@noop {} {\bibfield  {journal} {\bibinfo  {journal}
  {Science}\ }\textbf {\bibinfo {volume} {314}},\ \bibinfo {pages} {1757}
  (\bibinfo {year} {2006})}\BibitemShut {NoStop}%
\bibitem [{\citenamefont {Bernevig}\ and\ \citenamefont
  {Zhang}(2006)}]{BernevigPRL06}%
  \BibitemOpen
  \bibfield  {author} {\bibinfo {author} {\bibfnamefont {B.~A.}\ \bibnamefont
  {Bernevig}}\ and\ \bibinfo {author} {\bibfnamefont {S.-C.}\ \bibnamefont
  {Zhang}},\ }\href@noop {} {\bibfield  {journal} {\bibinfo  {journal} {Phys.
  Rev. Lett.}\ }\textbf {\bibinfo {volume} {96}},\ \bibinfo {pages} {106802}
  (\bibinfo {year} {2006})}\BibitemShut {NoStop}%
\bibitem [{\citenamefont {K{\"{o}}nig}\ \emph {et~al.}(2008)\citenamefont
  {K{\"{o}}nig}, \citenamefont {Buhmann}, \citenamefont {Molenkamp},
  \citenamefont {Hughes}, \citenamefont {Liu}, \citenamefont {Qi},\ and\
  \citenamefont {Zhang}}]{KonigJPSJ08}%
  \BibitemOpen
  \bibfield  {author} {\bibinfo {author} {\bibfnamefont {M.}~\bibnamefont
  {K{\"{o}}nig}}, \bibinfo {author} {\bibfnamefont {H.}~\bibnamefont
  {Buhmann}}, \bibinfo {author} {\bibfnamefont {L.~W.}\ \bibnamefont
  {Molenkamp}}, \bibinfo {author} {\bibfnamefont {T.}~\bibnamefont {Hughes}},
  \bibinfo {author} {\bibfnamefont {C.-X.}\ \bibnamefont {Liu}}, \bibinfo
  {author} {\bibfnamefont {X.-L.}\ \bibnamefont {Qi}}, \ and\ \bibinfo {author}
  {\bibfnamefont {S.-C.}\ \bibnamefont {Zhang}},\ }\href@noop {} {\bibfield
  {journal} {\bibinfo  {journal} {J. Phys. Soc. Jpn.}\ }\textbf {\bibinfo
  {volume} {77}},\ \bibinfo {pages} {031007} (\bibinfo {year}
  {2008})}\BibitemShut {NoStop}%
\bibitem [{\citenamefont {Moore}(2010)}]{MooreNAT10}%
  \BibitemOpen
  \bibfield  {author} {\bibinfo {author} {\bibfnamefont {J.~E.}\ \bibnamefont
  {Moore}},\ }\href@noop {} {\bibfield  {journal} {\bibinfo  {journal}
  {Nature}\ }\textbf {\bibinfo {volume} {464}},\ \bibinfo {pages} {194}
  (\bibinfo {year} {2010})}\BibitemShut {NoStop}%
\bibitem [{\citenamefont {Hasan}\ and\ \citenamefont
  {Kane}(2010)}]{HasanRMP10}%
  \BibitemOpen
  \bibfield  {author} {\bibinfo {author} {\bibfnamefont {M.~Z.}\ \bibnamefont
  {Hasan}}\ and\ \bibinfo {author} {\bibfnamefont {C.~L.}\ \bibnamefont
  {Kane}},\ }\href@noop {} {\bibfield  {journal} {\bibinfo  {journal} {Rev.
  Mod. Phys.}\ }\textbf {\bibinfo {volume} {82}},\ \bibinfo {pages} {3045}
  (\bibinfo {year} {2010})}\BibitemShut {NoStop}%
\bibitem [{\citenamefont {Hsieh}\ \emph {et~al.}(2008)\citenamefont {Hsieh},
  \citenamefont {Qian}, \citenamefont {Wray}, \citenamefont {Xia},
  \citenamefont {Hor}, \citenamefont {Cava},\ and\ \citenamefont
  {Hasan}}]{HsiehNAT08}%
  \BibitemOpen
  \bibfield  {author} {\bibinfo {author} {\bibfnamefont {D.}~\bibnamefont
  {Hsieh}}, \bibinfo {author} {\bibfnamefont {D.}~\bibnamefont {Qian}},
  \bibinfo {author} {\bibfnamefont {L.}~\bibnamefont {Wray}}, \bibinfo {author}
  {\bibfnamefont {Y.}~\bibnamefont {Xia}}, \bibinfo {author} {\bibfnamefont
  {Y.~S.}\ \bibnamefont {Hor}}, \bibinfo {author} {\bibfnamefont {R.~J.}\
  \bibnamefont {Cava}}, \ and\ \bibinfo {author} {\bibfnamefont {M.~Z.}\
  \bibnamefont {Hasan}},\ }\href@noop {} {\bibfield  {journal} {\bibinfo
  {journal} {Nature}\ }\textbf {\bibinfo {volume} {452}},\ \bibinfo {pages}
  {970} (\bibinfo {year} {2008})}\BibitemShut {NoStop}%
\bibitem [{\citenamefont {Hor}\ \emph {et~al.}(2009)\citenamefont {Hor},
  \citenamefont {Richardella}, \citenamefont {Roushan}, \citenamefont {Xia},
  \citenamefont {Checkelsky}, \citenamefont {Yazdani}, \citenamefont {Hasan},
  \citenamefont {Ong},\ and\ \citenamefont {Cava}}]{HorPRB09}%
  \BibitemOpen
  \bibfield  {author} {\bibinfo {author} {\bibfnamefont {Y.~S.}\ \bibnamefont
  {Hor}}, \bibinfo {author} {\bibfnamefont {A.}~\bibnamefont {Richardella}},
  \bibinfo {author} {\bibfnamefont {P.}~\bibnamefont {Roushan}}, \bibinfo
  {author} {\bibfnamefont {Y.}~\bibnamefont {Xia}}, \bibinfo {author}
  {\bibfnamefont {J.~G.}\ \bibnamefont {Checkelsky}}, \bibinfo {author}
  {\bibfnamefont {A.}~\bibnamefont {Yazdani}}, \bibinfo {author} {\bibfnamefont
  {M.~Z.}\ \bibnamefont {Hasan}}, \bibinfo {author} {\bibfnamefont {N.~P.}\
  \bibnamefont {Ong}}, \ and\ \bibinfo {author} {\bibfnamefont {R.~J.}\
  \bibnamefont {Cava}},\ }\href@noop {} {\bibfield  {journal} {\bibinfo
  {journal} {Phys. Rev. B}\ }\textbf {\bibinfo {volume} {79}},\ \bibinfo
  {pages} {195208} (\bibinfo {year} {2009})}\BibitemShut {NoStop}%
\bibitem [{\citenamefont {Hsieh}\ \emph
  {et~al.}(2009{\natexlab{a}})\citenamefont {Hsieh}, \citenamefont {Xia},
  \citenamefont {Qian}, \citenamefont {Wray}, \citenamefont {Dil},
  \citenamefont {Meier}, \citenamefont {Osterwalder}, \citenamefont {Patthey},
  \citenamefont {Checkelsky}, \citenamefont {Ong}, \citenamefont {Fedorov},
  \citenamefont {Lin}, \citenamefont {Bansil}, \citenamefont {Grauer},
  \citenamefont {Hor}, \citenamefont {Cava},\ and\ \citenamefont
  {Hasan}}]{HsiehNAT09}%
  \BibitemOpen
  \bibfield  {author} {\bibinfo {author} {\bibfnamefont {D.}~\bibnamefont
  {Hsieh}}, \bibinfo {author} {\bibfnamefont {Y.}~\bibnamefont {Xia}}, \bibinfo
  {author} {\bibfnamefont {D.}~\bibnamefont {Qian}}, \bibinfo {author}
  {\bibfnamefont {L.}~\bibnamefont {Wray}}, \bibinfo {author} {\bibfnamefont
  {J.~H.}\ \bibnamefont {Dil}}, \bibinfo {author} {\bibfnamefont
  {F.}~\bibnamefont {Meier}}, \bibinfo {author} {\bibfnamefont
  {J.}~\bibnamefont {Osterwalder}}, \bibinfo {author} {\bibfnamefont
  {L.}~\bibnamefont {Patthey}}, \bibinfo {author} {\bibfnamefont {J.~G.}\
  \bibnamefont {Checkelsky}}, \bibinfo {author} {\bibfnamefont {N.~P.}\
  \bibnamefont {Ong}}, \bibinfo {author} {\bibfnamefont {A.~V.}\ \bibnamefont
  {Fedorov}}, \bibinfo {author} {\bibfnamefont {H.}~\bibnamefont {Lin}},
  \bibinfo {author} {\bibfnamefont {A.}~\bibnamefont {Bansil}}, \bibinfo
  {author} {\bibfnamefont {D.}~\bibnamefont {Grauer}}, \bibinfo {author}
  {\bibfnamefont {Y.~S.}\ \bibnamefont {Hor}}, \bibinfo {author} {\bibfnamefont
  {R.~J.}\ \bibnamefont {Cava}}, \ and\ \bibinfo {author} {\bibfnamefont
  {M.~Z.}\ \bibnamefont {Hasan}},\ }\href@noop {} {\bibfield  {journal}
  {\bibinfo  {journal} {Nature}\ }\textbf {\bibinfo {volume} {460}},\ \bibinfo
  {pages} {1101} (\bibinfo {year} {2009}{\natexlab{a}})}\BibitemShut {NoStop}%
\bibitem [{\citenamefont {Chen}\ \emph {et~al.}(2009)\citenamefont {Chen},
  \citenamefont {Analytis}, \citenamefont {Chu}, \citenamefont {Liu},
  \citenamefont {Mo}, \citenamefont {Qi}, \citenamefont {Zhang}, \citenamefont
  {Lu}, \citenamefont {Dai}, \citenamefont {Fang}, \citenamefont {Zhang},
  \citenamefont {Fisher}, \citenamefont {Hussain},\ and\ \citenamefont
  {Shen}}]{ChenSCI09}%
  \BibitemOpen
  \bibfield  {author} {\bibinfo {author} {\bibfnamefont {Y.~L.}\ \bibnamefont
  {Chen}}, \bibinfo {author} {\bibfnamefont {J.~G.}\ \bibnamefont {Analytis}},
  \bibinfo {author} {\bibfnamefont {J.-H.}\ \bibnamefont {Chu}}, \bibinfo
  {author} {\bibfnamefont {Z.~K.}\ \bibnamefont {Liu}}, \bibinfo {author}
  {\bibfnamefont {S.-K.}\ \bibnamefont {Mo}}, \bibinfo {author} {\bibfnamefont
  {X.~L.}\ \bibnamefont {Qi}}, \bibinfo {author} {\bibfnamefont {H.~J.}\
  \bibnamefont {Zhang}}, \bibinfo {author} {\bibfnamefont {D.~H.}\ \bibnamefont
  {Lu}}, \bibinfo {author} {\bibfnamefont {X.}~\bibnamefont {Dai}}, \bibinfo
  {author} {\bibfnamefont {Z.}~\bibnamefont {Fang}}, \bibinfo {author}
  {\bibfnamefont {S.~C.}\ \bibnamefont {Zhang}}, \bibinfo {author}
  {\bibfnamefont {I.~R.}\ \bibnamefont {Fisher}}, \bibinfo {author}
  {\bibfnamefont {Z.}~\bibnamefont {Hussain}}, \ and\ \bibinfo {author}
  {\bibfnamefont {Z.-X.}\ \bibnamefont {Shen}},\ }\href@noop {} {\bibfield
  {journal} {\bibinfo  {journal} {Science}\ }\textbf {\bibinfo {volume}
  {325}},\ \bibinfo {pages} {178} (\bibinfo {year} {2009})}\BibitemShut
  {NoStop}%
\bibitem [{\citenamefont {Hsieh}\ \emph
  {et~al.}(2009{\natexlab{b}})\citenamefont {Hsieh}, \citenamefont {Xia},
  \citenamefont {Qian}, \citenamefont {Wray}, \citenamefont {Meier},
  \citenamefont {Dil}, \citenamefont {Osterwalder}, \citenamefont {Patthey},
  \citenamefont {Fedorov}, \citenamefont {Lin}, \citenamefont {Bansil},
  \citenamefont {Grauer}, \citenamefont {Hor}, \citenamefont {Cava},\ and\
  \citenamefont {Hasan}}]{HsiehPRL09}%
  \BibitemOpen
  \bibfield  {author} {\bibinfo {author} {\bibfnamefont {D.}~\bibnamefont
  {Hsieh}}, \bibinfo {author} {\bibfnamefont {Y.}~\bibnamefont {Xia}}, \bibinfo
  {author} {\bibfnamefont {D.}~\bibnamefont {Qian}}, \bibinfo {author}
  {\bibfnamefont {L.}~\bibnamefont {Wray}}, \bibinfo {author} {\bibfnamefont
  {F.}~\bibnamefont {Meier}}, \bibinfo {author} {\bibfnamefont {J.~H.}\
  \bibnamefont {Dil}}, \bibinfo {author} {\bibfnamefont {J.}~\bibnamefont
  {Osterwalder}}, \bibinfo {author} {\bibfnamefont {L.}~\bibnamefont
  {Patthey}}, \bibinfo {author} {\bibfnamefont {A.~V.}\ \bibnamefont
  {Fedorov}}, \bibinfo {author} {\bibfnamefont {H.}~\bibnamefont {Lin}},
  \bibinfo {author} {\bibfnamefont {A.}~\bibnamefont {Bansil}}, \bibinfo
  {author} {\bibfnamefont {D.}~\bibnamefont {Grauer}}, \bibinfo {author}
  {\bibfnamefont {Y.~S.}\ \bibnamefont {Hor}}, \bibinfo {author} {\bibfnamefont
  {R.~J.}\ \bibnamefont {Cava}}, \ and\ \bibinfo {author} {\bibfnamefont
  {M.~Z.}\ \bibnamefont {Hasan}},\ }\href@noop {} {\bibfield  {journal}
  {\bibinfo  {journal} {Phys. Rev. Lett.}\ }\textbf {\bibinfo {volume} {103}},\
  \bibinfo {pages} {146401} (\bibinfo {year} {2009}{\natexlab{b}})}\BibitemShut
  {NoStop}%
\bibitem [{\citenamefont {Chadov}\ \emph {et~al.}(2010)\citenamefont {Chadov},
  \citenamefont {Qi}, \citenamefont {K\'{u}bler}, \citenamefont {Fecher},
  \citenamefont {Fecher},\ and\ \citenamefont {Zhang}}]{ChadovNMAT10}%
  \BibitemOpen
  \bibfield  {author} {\bibinfo {author} {\bibfnamefont {S.}~\bibnamefont
  {Chadov}}, \bibinfo {author} {\bibfnamefont {X.}~\bibnamefont {Qi}}, \bibinfo
  {author} {\bibfnamefont {J.}~\bibnamefont {K\'{u}bler}}, \bibinfo {author}
  {\bibfnamefont {G.~H.}\ \bibnamefont {Fecher}}, \bibinfo {author}
  {\bibfnamefont {C.}~\bibnamefont {Fecher}}, \ and\ \bibinfo {author}
  {\bibfnamefont {S.~C.}\ \bibnamefont {Zhang}},\ }\href@noop {} {\bibfield
  {journal} {\bibinfo  {journal} {Nature Mat.}\ }\textbf {\bibinfo {volume}
  {9}},\ \bibinfo {pages} {541} (\bibinfo {year} {2010})}\BibitemShut {NoStop}%
\bibitem [{\citenamefont {Lin}\ \emph {et~al.}(2010)\citenamefont {Lin},
  \citenamefont {Wray}, \citenamefont {Xia}, \citenamefont {Xu}, \citenamefont
  {Jia}, \citenamefont {Cava}, \citenamefont {Bansil},\ and\ \citenamefont
  {Hasan}}]{LinNMAT10}%
  \BibitemOpen
  \bibfield  {author} {\bibinfo {author} {\bibfnamefont {H.}~\bibnamefont
  {Lin}}, \bibinfo {author} {\bibfnamefont {L.~A.}\ \bibnamefont {Wray}},
  \bibinfo {author} {\bibfnamefont {Y.}~\bibnamefont {Xia}}, \bibinfo {author}
  {\bibfnamefont {S.}~\bibnamefont {Xu}}, \bibinfo {author} {\bibfnamefont
  {S.}~\bibnamefont {Jia}}, \bibinfo {author} {\bibfnamefont {R.~J.}\
  \bibnamefont {Cava}}, \bibinfo {author} {\bibfnamefont {A.}~\bibnamefont
  {Bansil}}, \ and\ \bibinfo {author} {\bibfnamefont {M.~Z.}\ \bibnamefont
  {Hasan}},\ }\href@noop {} {\bibfield  {journal} {\bibinfo  {journal} {Nature
  Mat.}\ }\textbf {\bibinfo {volume} {9}},\ \bibinfo {pages} {546} (\bibinfo
  {year} {2010})}\BibitemShut {NoStop}%
\bibitem [{\citenamefont {Pesin}\ and\ \citenamefont
  {Balents}(2010)}]{PesinNPHY10}%
  \BibitemOpen
  \bibfield  {author} {\bibinfo {author} {\bibfnamefont {D.}~\bibnamefont
  {Pesin}}\ and\ \bibinfo {author} {\bibfnamefont {L.}~\bibnamefont
  {Balents}},\ }\href@noop {} {\bibfield  {journal} {\bibinfo  {journal}
  {Nature Phys.}\ }\textbf {\bibinfo {volume} {6}},\ \bibinfo {pages} {376}
  (\bibinfo {year} {2010})}\BibitemShut {NoStop}%
\bibitem [{\citenamefont {Rachel}\ and\ \citenamefont
  {Hur}(2010)}]{RachelPRB10}%
  \BibitemOpen
  \bibfield  {author} {\bibinfo {author} {\bibfnamefont {S.}~\bibnamefont
  {Rachel}}\ and\ \bibinfo {author} {\bibfnamefont {K.~L.}\ \bibnamefont
  {Hur}},\ }\href@noop {} {\bibfield  {journal} {\bibinfo  {journal} {Phys.
  Rev. B}\ }\textbf {\bibinfo {volume} {82}},\ \bibinfo {pages} {075106}
  (\bibinfo {year} {2010})}\BibitemShut {NoStop}%
\bibitem [{\citenamefont {Varney}\ \emph {et~al.}(2010)\citenamefont {Varney},
  \citenamefont {Sun}, \citenamefont {Rigol},\ and\ \citenamefont
  {Galitski}}]{VarneyPRB10}%
  \BibitemOpen
  \bibfield  {author} {\bibinfo {author} {\bibfnamefont {C.~N.}\ \bibnamefont
  {Varney}}, \bibinfo {author} {\bibfnamefont {K.}~\bibnamefont {Sun}},
  \bibinfo {author} {\bibfnamefont {M.}~\bibnamefont {Rigol}}, \ and\ \bibinfo
  {author} {\bibfnamefont {V.}~\bibnamefont {Galitski}},\ }\href@noop {}
  {\bibfield  {journal} {\bibinfo  {journal} {Phys. Rev. B}\ }\textbf {\bibinfo
  {volume} {82}},\ \bibinfo {pages} {115125} (\bibinfo {year}
  {2010})}\BibitemShut {NoStop}%
\bibitem [{\citenamefont {Yu}\ \emph {et~al.}(2011)\citenamefont {Yu},
  \citenamefont {Xie},\ and\ \citenamefont {Li}}]{YuPRL11}%
  \BibitemOpen
  \bibfield  {author} {\bibinfo {author} {\bibfnamefont {S.-L.}\ \bibnamefont
  {Yu}}, \bibinfo {author} {\bibfnamefont {X.~C.}\ \bibnamefont {Xie}}, \ and\
  \bibinfo {author} {\bibfnamefont {J.-X.}\ \bibnamefont {Li}},\ }\href@noop {}
  {\bibfield  {journal} {\bibinfo  {journal} {Phys. Rev. Lett.}\ }\textbf
  {\bibinfo {volume} {107}},\ \bibinfo {pages} {010410} (\bibinfo {year}
  {2011})}\BibitemShut {NoStop}%
\bibitem [{\citenamefont {He}\ \emph {et~al.}(2011)\citenamefont {He},
  \citenamefont {Kou}, \citenamefont {Liang},\ and\ \citenamefont
  {Feng}}]{HePRB11}%
  \BibitemOpen
  \bibfield  {author} {\bibinfo {author} {\bibfnamefont {J.}~\bibnamefont
  {He}}, \bibinfo {author} {\bibfnamefont {S.-P.}\ \bibnamefont {Kou}},
  \bibinfo {author} {\bibfnamefont {Y.}~\bibnamefont {Liang}}, \ and\ \bibinfo
  {author} {\bibfnamefont {S.}~\bibnamefont {Feng}},\ }\href@noop {} {\bibfield
   {journal} {\bibinfo  {journal} {Phys. Rev. B}\ }\textbf {\bibinfo {volume}
  {83}},\ \bibinfo {pages} {205116} (\bibinfo {year} {2011})}\BibitemShut
  {NoStop}%
\bibitem [{\citenamefont {Yamaji}\ and\ \citenamefont
  {Imada}(2011)}]{YamajiPRB11}%
  \BibitemOpen
  \bibfield  {author} {\bibinfo {author} {\bibfnamefont {Y.}~\bibnamefont
  {Yamaji}}\ and\ \bibinfo {author} {\bibfnamefont {M.}~\bibnamefont {Imada}},\
  }\href@noop {} {\bibfield  {journal} {\bibinfo  {journal} {Phys. Rev. B}\
  }\textbf {\bibinfo {volume} {83}},\ \bibinfo {pages} {205122} (\bibinfo
  {year} {2011})}\BibitemShut {NoStop}%
\bibitem [{\citenamefont {Li}\ \emph {et~al.}(2009)\citenamefont {Li},
  \citenamefont {Chu}, \citenamefont {Jain},\ and\ \citenamefont
  {Shen}}]{LiPRL09}%
  \BibitemOpen
  \bibfield  {author} {\bibinfo {author} {\bibfnamefont {J.}~\bibnamefont
  {Li}}, \bibinfo {author} {\bibfnamefont {R.-L.}\ \bibnamefont {Chu}},
  \bibinfo {author} {\bibfnamefont {J.~K.}\ \bibnamefont {Jain}}, \ and\
  \bibinfo {author} {\bibfnamefont {S.-Q.}\ \bibnamefont {Shen}},\ }\href
  {\doibase 10.1103/PhysRevLett.102.136806} {\bibfield  {journal} {\bibinfo
  {journal} {Phys. Rev. Lett.}\ }\textbf {\bibinfo {volume} {102}},\ \bibinfo
  {pages} {136806} (\bibinfo {year} {2009})}\BibitemShut {NoStop}%
\bibitem [{\citenamefont {Brinkman}\ and\ \citenamefont
  {Rice}(1970)}]{Brinkman1970}%
  \BibitemOpen
  \bibfield  {author} {\bibinfo {author} {\bibfnamefont {W.~F.}\ \bibnamefont
  {Brinkman}}\ and\ \bibinfo {author} {\bibfnamefont {T.~M.}\ \bibnamefont
  {Rice}},\ }\href {\doibase 10.1103/PhysRevB.2.4302} {\bibfield  {journal}
  {\bibinfo  {journal} {Phys. Rev. B}\ }\textbf {\bibinfo {volume} {2}},\
  \bibinfo {pages} {4302} (\bibinfo {year} {1970})}\BibitemShut {NoStop}%
\bibitem [{\citenamefont {Florens}\ and\ \citenamefont
  {Georges}(2002)}]{FlorensPRB02}%
  \BibitemOpen
  \bibfield  {author} {\bibinfo {author} {\bibfnamefont {S.}~\bibnamefont
  {Florens}}\ and\ \bibinfo {author} {\bibfnamefont {A.}~\bibnamefont
  {Georges}},\ }\href@noop {} {\bibfield  {journal} {\bibinfo  {journal} {Phys.
  Rev. B}\ }\textbf {\bibinfo {volume} {66}},\ \bibinfo {pages} {165111}
  (\bibinfo {year} {2002})}\BibitemShut {NoStop}%
\bibitem [{\citenamefont {Florens}\ and\ \citenamefont
  {Georges}(2004)}]{FlorensPRB04}%
  \BibitemOpen
  \bibfield  {author} {\bibinfo {author} {\bibfnamefont {S.}~\bibnamefont
  {Florens}}\ and\ \bibinfo {author} {\bibfnamefont {A.}~\bibnamefont
  {Georges}},\ }\href@noop {} {\bibfield  {journal} {\bibinfo  {journal} {Phys.
  Rev. B}\ }\textbf {\bibinfo {volume} {70}},\ \bibinfo {pages} {035114}
  (\bibinfo {year} {2004})}\BibitemShut {NoStop}%
\bibitem [{\citenamefont {Zhao}\ and\ \citenamefont
  {Paramekanti}(2007)}]{ZhaoPRB07}%
  \BibitemOpen
  \bibfield  {author} {\bibinfo {author} {\bibfnamefont {E.}~\bibnamefont
  {Zhao}}\ and\ \bibinfo {author} {\bibfnamefont {A.}~\bibnamefont
  {Paramekanti}},\ }\href@noop {} {\bibfield  {journal} {\bibinfo  {journal}
  {Phys. Rev. B}\ }\textbf {\bibinfo {volume} {76}},\ \bibinfo {pages} {195101}
  (\bibinfo {year} {2007})}\BibitemShut {NoStop}%
\bibitem [{\citenamefont {Fu}\ and\ \citenamefont {Kane}(2007)}]{FuPRB07}%
  \BibitemOpen
  \bibfield  {author} {\bibinfo {author} {\bibfnamefont {L.}~\bibnamefont
  {Fu}}\ and\ \bibinfo {author} {\bibfnamefont {C.~L.}\ \bibnamefont {Kane}},\
  }\href@noop {} {\bibfield  {journal} {\bibinfo  {journal} {Phys. Rev. B}\
  }\textbf {\bibinfo {volume} {76}},\ \bibinfo {pages} {045302} (\bibinfo
  {year} {2007})}\BibitemShut {NoStop}%
\bibitem [{Note1()}]{Note1}%
  \BibitemOpen
  \bibinfo {note} {Since our focus is on the charge sector and the associated
  Mott physics in the spirit of \cite {Brinkman1970}, we keep aside the
  magnetic correlations which are typically introduced by adding an exchange
  term in the Hamiltonian\cite {RachelPRB10} and treating it via mean field
  theory.}\BibitemShut {Stop}%
\end{thebibliography}%
\end{document}